\documentclass[useAMS,usenatbib]{mn2e}
\usepackage{psfig}
\usepackage{graphicx}
\usepackage{times}
\usepackage{color}                % for colour font
\def\kms{km ${\rm s}^{-1}$}

\def\ch2{$\chi^2$}
\def\dg{$^{\circ}$}
\def\kms {\hbox{${\rm km\ s}^{-1}$}}

\def\scm  {$\hbox{{\rm cm}}^{-2}$}    %cm-2
\def\MOLH {\hbox{${\rm H}_2$}}  %H2
\def \AL {$\alpha $}     %  gr. alpha
\def \HI {H{\sc \,i}}
\def \WpHz {W Hz$^{-1}$}

\def\lapp{\ifmmode\stackrel{<}{_{\sim}}\else$\stackrel{<}{_{\sim}}$\fi}
\def\gapp{\ifmmode\stackrel{>}{_{\sim}}\else$\stackrel{>}{_{\sim}}$\fi}

\title[High redshift millimetre absorption]{On the absence of molecular absorption in high redshift
  millimetre-band searches}

 \author[S. J. Curran et al.]{S. J. Curran$^{1}$\thanks{E-mail: sjc@phys.unsw.edu.au}, M. T. Whiting$^{1,2}$, F. Combes$^{3}$, N. Kuno$^{4}$, P. Francis$^{5}$, 
N. Nakai$^{6}$, \newauthor J. K. Webb$^{1}$, M. T. Murphy$^{1,7}$ and T. Wiklind$^{8,9,10}$\\
$^{1}$School of Physics, University of New South Wales, Sydney NSW 2052, Australia\\
$^{2}$CSIRO Australia Telescope National Facility, PO Box 76, Epping NSW 1710, Australia\\
$^{3}$LERMA, Observatoire de Paris, France\\
$^{4}$Nobeyama Radio Observatory, Nagano 384-1305, Japan\\
$^{5}$Australian National University, Australia\\ 
$^{6}$Institute of Physics, University of Tsukuba, Ten-noudai, Tsukuba, Ibaraki 
305-8571, Japan\\
$^{7}$Centre for Astrophysics and Supercomputing, Swinburne University of Technology, PO Box 218, Hawthorn, VIC 3122, Australia\\
$^{8}$Space Telescope Science Institute, USA\\
$^{9}$Onsala Space Observatory, S-439 92 Onsala, Sweden\\
$^{10}$Joint ALMA Observatory, Santiago, Chile}
\begin{document}

\date{Accepted ---. Received ---; in original form ---}

\pagerange{\pageref{firstpage}--\pageref{lastpage}} \pubyear{2011}

\maketitle

\label{firstpage}

\begin{abstract}
  We have undertaken a search for millimetre-wave band absorption (through the CO and HCO$^+$ rotational transitions) in
  the host galaxies of reddened radio sources ($z = 0.405-1.802$). Despite the colour selection (optical--near infrared
  colours of $V - K\gapp5$ in all but one source), no absorption was found in any of the eight quasars for which the
  background continuum flux was detected.  On the basis of the previous (mostly intervening) \MOLH\ and OH detections,
  the limits reached here and in some previous surveys should be deep enough to detect molecular absorption according to
  their $V - K$ colours. However, our survey makes the assumption that the reddening is associated with dust close
to the emission redshift of the quasar and that the narrow millimetre component of this emission is intercepted by the compact
molecular cores. By using the known millimetre absorbers to define the {\em colour depth} and comparing this with the ultra-violet
  luminosities of the sources, we find that, even {\em if} these assumptions are valid, only twelve of the forty objects
  (mainly from this work) are potentially detectable. This is assuming an excitation temperature of $T_{\rm x} = 10$ K
  at $z=0$, with the number decreasing with increasing temperatures (to zero detectable at $T_{\rm x} \gapp100$ K).
\end{abstract}


\section{Introduction}
\label{sec:intro}

Millimetre-wave observations of molecular absorption systems along the sight-lines to distant
quasars provide a powerful probe of the cold, dense, star forming gas in the distant Universe.
Furthermore, comparison of the redshifts of the rotational transitions of the molecules with those of the spin-flip transition
of \HI, as well as the  electronic optical/UV transitions of metal ions, can be used to determine high redshift values
of the fundamental constants, to at least an order of magnitude the sensitivity of purely optical
data (see \citealt{cdk04}).
However, despite
much searching, only four such systems are currently known \citep{wc95,wc96,wc96b,wc97}, the highest
redshift being at $z_{\rm abs} = 0.89$. Of these, two are intervening systems (gravitational lenses
towards more distant quasars), with the other two systems arising through absorption within the host
galaxy of the quasar. Subsequent searches at the redshifts of known high column density \HI\
absorption systems, intervening the sight-lines to more distant quasi-stellar objects (QSOs), have also
failed to detect molecular absorption in the millimetre-band (\citealt{cmpw03} and references
therein), despite the possibility that these so-called damped Lyman-\AL\ systems (DLAs)\footnote{These have neutral
  hydrogen column densities of $N_{\rm HI}\geq2\times10^{20}$ \scm\ and are usually detected at
  $z_{\rm abs}\gapp1.8$, where the Lyman-\AL\ transition is redshifted in to the optical band.} may
account for more than 80 per-cent of the neutral gas content in the Universe \citep{phw05}.

\begin{figure}
\centering \includegraphics[angle=270,scale=0.43]{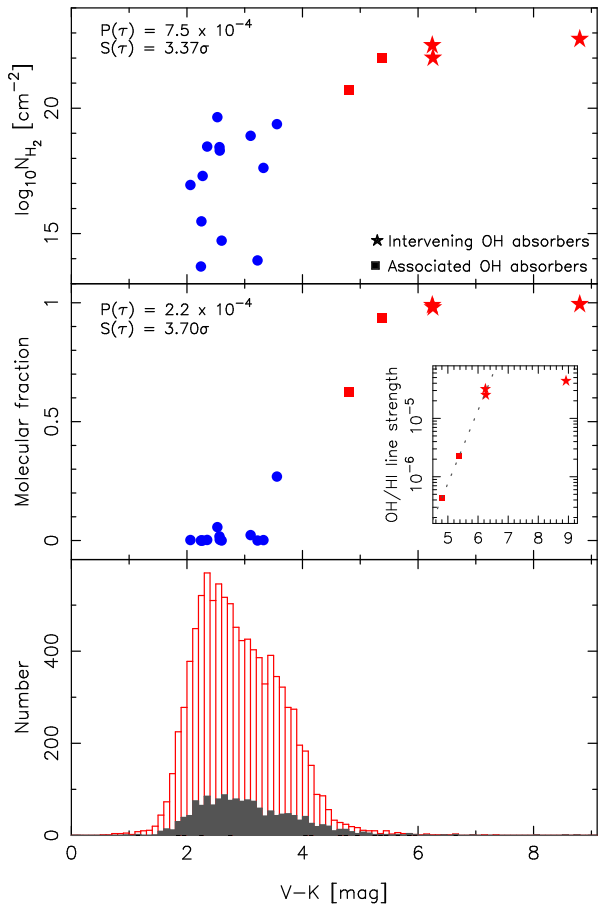} %$
\caption{The \MOLH\ column density (top) and the molecular fraction (middle) 
versus the  observed frame optical--near infrared colour (where available) for high redshift
  molecular absorption systems. The circles represent the
  \MOLH-bearing DLAs (all optically selected intervening absorbers)
  and the squares and stars the OH absorbers (radio selected), with the inset in the middle panel
showing the normalised OH line strength \citep{cwm+06}. 
These
  are comprised of the four systems originally identified in
  millimetre-wave transitions 
(with the least-squares fit to these shown)
plus the gravitational lens at $z_{\rm abs} =0.764$ towards
  0132--097 (detected in OH decimetre but not HCO$^+$ millimetre absorption,
  \citealt{kcl+05}). 
$P(\tau)$ shows Kendall's $\tau$ two-sided probability of the observed distribution
occuring by chance and $S(\tau)$ the significance of this assuming Gaussian statistics.
The bottom panel shows the distribution of SDSS quasars (\protect\citealt{shr+07}, after conversion to $V$ magnitudes, \citealt{fig+96})
 where the filled histogram shows those which have been detected in the  Very
Large Array's ``Faint Images of the Radio Sky at Twenty
Centimetres'' (FIRST) survey \protect\citep{bwh95}.}
\label{frac-col}
\end{figure} 
DLAs are, however, not devoid of molecular gas: To date, the Lyman and Werner ultra-violet bands of \MOLH\ have been
detected in 19 DLAs (see \citealt{nlps08}\footnote{One of which, J1439+113, has also been detected in the CO $A-X$ UV
  band \citep{snlp08}.}, \citealt{jwpc09} and \citealt{sgp+10}). These, however, have molecular abundances which are
generally much lower than those detectable with current microwave and radio telescopes (\citealt{cmpw03} and
Fig. \ref{frac-col}, top).  Furthermore, in \citet{cwm+06} we showed that the \MOLH-bearing DLAs have 
molecular fractions of
${\cal  F}\equiv\frac{2N_{\rm H_2}}{2N_{\rm H_2}+N_{\rm HI}}\sim10^{-7} - 0.3$ and $V-K\lapp4$ (Fig.~\ref{frac-col}, middle),
i.e. in the same range as a "typical" QSO (Fig. \ref{frac-col}, bottom)\footnote{$V-K = 2.88\pm1.04$ in general and
  $3.05\pm0.97$ if radio-loud.}, whereas the millimetre and decimetre band absorbers have molecular fractions ${\cal
  F}\approx0.6 - 1$ and optical--near-infrared colours of $V-K\gapp5$.

The correlations in Fig.~\ref{frac-col}  present strong evidence that the quasar light is
reddened by dust in the foreground absorber: Since the presence of the dust
is necessary to prevent the dissociation of the molecular gas by the
ambient ultra-violet field, the molecular fraction is expected
to be correlated with the dust abundance, as observed.  The paucity of
millimetre-wave band absorption can therefore be attributed to the
traditional optical selection of targets biasing towards absorbers of low dust content and therefore low molecular
fractions. 

The fact that intervening absorbers are usually found through optical spectroscopy, yielding a redshift but also giving
the above bias against dusty objects, means that millimetre-band searches of known intervening absorbers have generally
been unsuccessful (\citealt{cmpw03} and references therein). An alternative target for molecular absorption is towards
the fainter ``red quasars'', where the red colour may indicate an intervening column of dust. However, due to the relatively
narrow bandwidths in the millimetre band (see Sect. \ref{lotag}), such an approach is currently only practical at longer (decimetre)
wavelengths (see \citealt{cwt+11}). In the absence of any known intervening absorbers, selecting the quasar
itself gives a redshift ($z_{\rm em}$) to which to tune the receiver. Naturally, such a selection of targets prevents any useful comparison 
with the optical redshifts, in order to measure the values of the fundamental constants, although any detections
could be followed up in 21-cm, giving the redshift of the spin-flip transition of \HI.

In \citet{cwm+06,cww+08,cwm+10}, we presented the results from
our decimetre-wave searches for such ``associated'' (OH and \HI) absorption and here present the results of our
 millimetre-wave survey for associated absorption.

\section{Observations}\label{obs}
%\subsection{Observations}
\subsection{Target selection}
\label{ts}

As per \citet{cwm+06}, our sources were selected from the Parkes Half-Jansky
Flat-spectrum Sample (PHFS, \citealt{dwf+97})\footnote{With the addition of  0500+019,
included since it has been detected in
21-cm absorption  \citep{cmr+98}. We also included  J0906+4952 \& J1341+3301, which are two
  very red sources from \citet{ggl+04} [Sect. \ref{nob}].}, on the
basis of their optical--near-IR photometry \citep{fww00}. From these,
we selected the 30 reddest sources (which correspond to an extinction
of $A_V\approx4.1$), in which the emission redshift of the quasar ($z_{\rm em}$) would place a
strong absorption line (CO or HCO$^+$) into the 3-mm band. After
culling these further, by selecting those of $\delta>-30$\dg\ (thus
being observable from northern latitudes)\footnote{We miss SEST.} and
with 3-mm flux densities expected to be $\gapp100$ mJy, the  ten
objects listed in Table~\ref{sum} remained.
 
\subsection{The IRAM 30-m observations}

From December 2003 to February 2004 we observed three of the targets
with the IRAM (Institut de Radio Astronomie Millimetrique) 30-m
telescope at Pico Veleta, Spain. We used two 3-mm SIS receivers (A100
and B100), tuned to the redshifted frequencies of the molecular
transitions (see Table \ref{sum}). The observations were done with a
nutating subreflector, switching symmetrically $\pm90''$ in azimuth
with a frequency of 0.5 Hz. The continuum levels of the observed
sources were determined using a continuum backend and increasing the
subreflector switch frequency to 2 Hz.  The image sideband rejection
of the receivers were high, of the order of 20 dB (single
sideband). System temperatures typically ranged from 120 K (90 GHz) to
180 K (113 GHz). We used the full 1 GHz backend with broad (1MHz)
filterbanks and narrow band autocorrelators, the former giving a
channel spacing  of $\approx3-4$ \kms\ (Table \ref{sum}) over a
bandwidth of $\approx1500$ \kms, which should be sufficient to
cover uncertainties in the emission redshifts, all of which are
known to at least three decimal places ($\Delta z = \pm0.001$ corresponds
to $\Delta v \approx \pm 100 - 200$ \kms\ for our sample). 
The pointing of the telescope was
checked regularly on nearby continuum sources. Typical pointing
corrections were $5-10''$.  The focus was checked regularly on Mars
and Saturn.  The Half Power Beam Width (HPBW) at 95 GHz is $26''$.

We prioritised the three targets according to the 3-mm flux densities
estimated from an interpolation of the decimetre and near-infrared
values: 12.8 hours of integration on 0500+019 ($S_{\rm est}\approx0.32$
Jy), 13.6 hours on 1430--155 ($S_{\rm est}\gapp0.08$ Jy) and 7.2 hours
on 1504--166 ($S_{\rm est}\gapp0.8$ Jy), which, upon comparison with
the observed values (Table \ref{sum}), were reasonable estimates. The
data were reduced with the {\sc gildas}\footnote{http://www.iram.fr/IRAMFR/GILDAS/} software
package.

\subsection{The NRO 45-m observations}
\label{nob}

The remainder of the sample was observed with the Nobeyama Radio
Observatory's 45-m telescope in March 2004. We used the H28/32 (1-cm)
and the S80, S100 (3-mm) receivers to observe the $J=0\rightarrow1$
and $1\rightarrow2$ transitions over a range of redshifts
(Table~\ref{sum}). The observations were performed in position
switching mode, with an integration time of 20 s for each scan. The
antenna temperature, $T_{\rm A}^*$ , was obtained by the chopper-wheel
method, which corrected for atmospheric and ohmic losses.  System
temperatures typically ranged from 200~K (1-cm) to 500~K (3-mm). We
used the AOS backend over 250 MHz split over 2048 channels, which gave
channel spacings of 0.37 (3-mm) \kms\ to 1.1 \kms\ (1-cm) with
bandwidths of $\approx700$ and $\approx2000$ \kms, respectively. The
pointing of the telescope was checked by observing SiO maser sources
with the H40 (40 GHz band) receiver and the corrections were
$\leq10"$.

The 3-mm observations were performed when the weather was clearest,
with total integration times of 2 hours for 0213--026, 5 hours for
0454+066, 3.7 hours for 1107--187 and 8.5 hours for 1706+006 (for
which we could not determine the flux density, Fig.~\ref{spectra})\footnote{The highest frequency flux measurement
  available for 1706+006  is 0.44 Jy at 5.0 GHz \citep{wo90}.}.  During less than
ideal weather conditions, we observed in the 1-cm band and included the radio
detected, optically dim sources of \citet{ggl+04},  where HCO$^+$
$0\rightarrow1$ is redshifted into this band --- the FIRST--2MASS
reddened quasars J0906+4952 (SDSS\,J090651.49+495235.9) and FTM\,J1341+3301. The
data were reduced with the {\sc newstar} package and, like the IRAM
results, upon the removal of a low order baseline and smoothing no
absorption features were apparent in the spectra.

%\section{Results and Discussion}
\section{Results}
\subsection{Observational results}
\label{obs}

\begin{figure*}
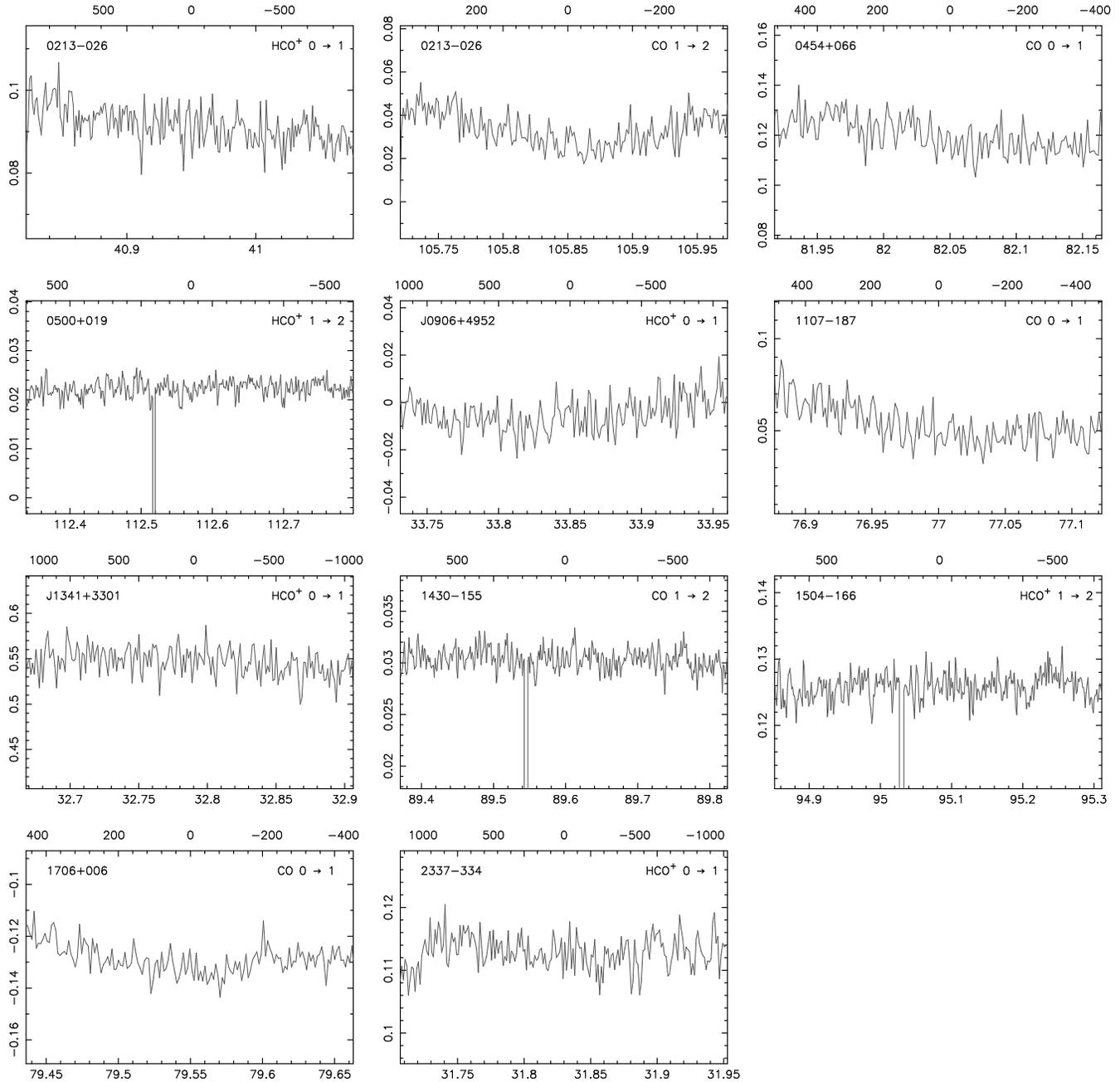

\vspace{17.4cm}
% were e.g. psfile= nob/0213-40-atsum.txt.ps  hoffset=-20 voffset=500 hscale=45 vscale=45 angle=-90} - saved in paper3_22March11.tex
% NEW ONES PRODUCED FROM spectrum-nro.c
\includegraphics{0213-40-atsum.txt.ps}
\includegraphics{0213-105-atsum.txt.ps}
\includegraphics{0454-atsum.txt.ps}
%new line  - from spectrum-iram.c
\includegraphics{0500+019.dat.ps}
\includegraphics{0906-atsum.txt.ps}
\includegraphics{1107-atsum.txt.ps}
%new line
\includegraphics{1341-atsum.txt.ps}
\includegraphics{1430-155.dat.ps}
\includegraphics{1504-166.dat.ps}
%new line
\includegraphics{1706-atsum.txt.ps}
\includegraphics{2337-sum-atsum.txt.ps}
%\special{psfile=  hoffset=320 voffset=140 hscale=45 vscale=45 angle=-90}
\caption{The observed spectra before baseline removal. The ordinate in each spectrum shows the antenna temperature in $T_{\rm A}^*$ [K] and
  the abscissa the barycentric frequency [GHz], with the top axis showing the velocity offset from the central tuned frequency [\kms].
All are shown at a resolution of 4~\kms, the coarsest of the observations (see Table \ref{sum}). The negative spike 
in each of the IRAM spectra are due to bad channels (246--250) and the negative flux for 1706+006 is due to the long (20 s)
integrations with the NRO giving errors in the continuum flux measurement. }
\label{spectra}
\end{figure*}
% putting in sidways table as a figure  
\begin{figure}
\vspace{26cm}
\includegraphics{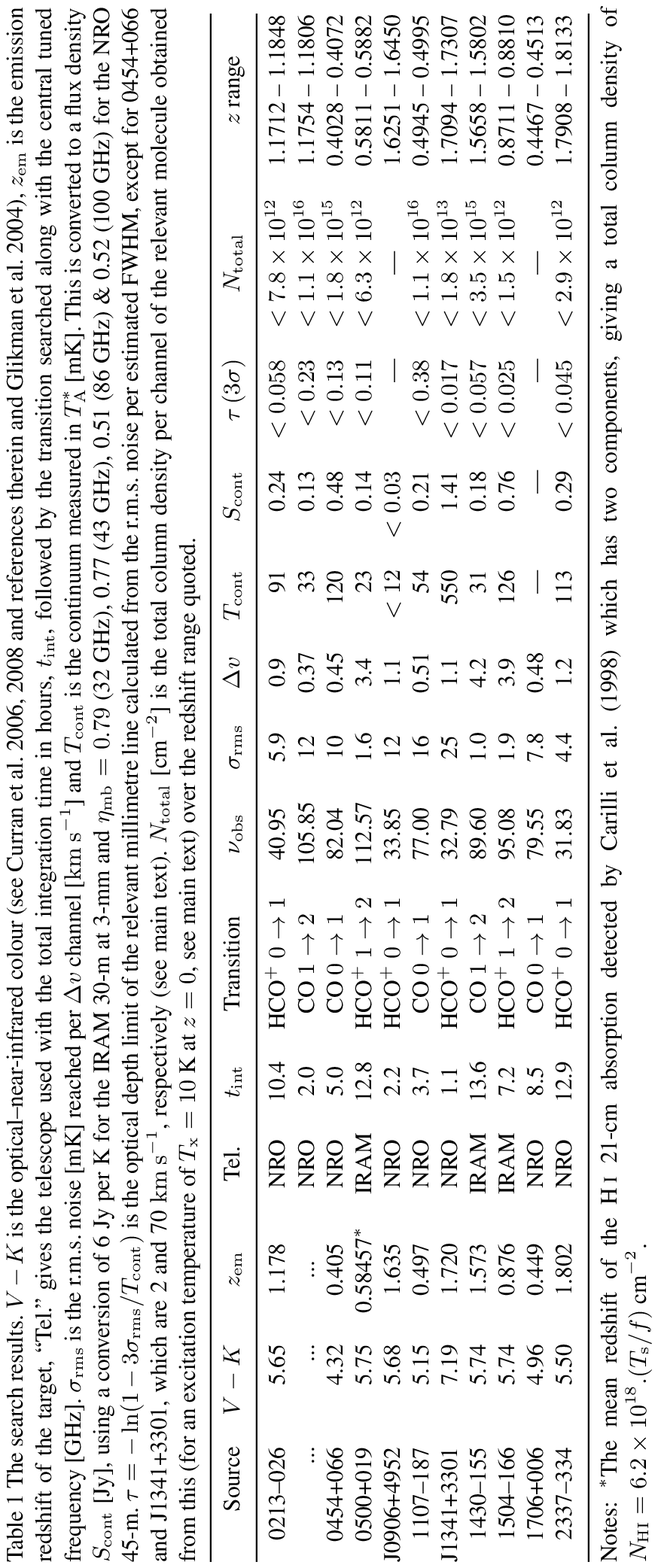}
\label{sum}
\addtocounter{table}{1}
\end{figure}
In Fig. \ref{spectra} we show the reduced spectra and in Table
\ref{sum} we summarise our observational results. The limit to
the total column density of each molecule is calculated from
\begin{equation}
 N_{\rm total}=\frac{8\pi}{c^3}\frac{\nu^{3}}{g_{J+1}A_{J+1\rightarrow J}}\frac{Qe^{E_J/kT_{\rm x}}}{1-e^{-h\nu/kT_{\rm x}}}
\left.\int\right.\tau dv,
\label{equ2}
\end{equation}
where $\nu$ is the rest frequency of the $J\rightarrow J+1$ transition,
$g_{J+1}$ and $A_{J+1\rightarrow J}$ are the statistical weight and
the Einstein A-coefficient of the transition, respectively, and $Q =
\sum^{\infty}_{J=0}g_{J}~e^{-E_J/kT_{\rm x}}$ is the partition
function\footnote{The Einstein A-coefficients are taken from
  \citet{cklh95,cms96} or derived from the permanent electric dipole
  moment of the molecule, obtained from the JPL Spectral
  Line Catalog \citep{ppc+98}, along with the energy of each level,
  $E_J$. An on-line column density calculator based on Equation \ref{equ2}
  is available from http://www.phys.unsw.edu.au/$\sim$sjc/column/}. In
the case of the four known systems, the covering factor is expected to
be close to unity \citep{wc94,wc95,wc96b,wc98} so, as in the optically
thin regime, this can be written outside of the integral. However, unlike
lower frequency searches (e.g. OH 18-cm, $^{2}\Pi_{3/2}$), in the
millimetre regime even low excitation temperatures give $kT_{\rm
  x}\sim E_J\sim h\nu$ and so the column density cannot be
approximated via a linear dependence on this. Therefore, unlike the
decimetre searches for associated absorption \citep{cwm+06,cww+08,cwm+10}, we must
assume an excitation temperature.

In Table \ref{columns} we apply Equation \ref{equ2} to the velocity
integrated optical depths given in the references in order to derive
the column densities and excitation temperatures of the four known
systems.
\begin{table*} 
\centering
\begin{minipage}{148mm}
\caption{The excitation temperatures  and column densities derived from the listed references.
In the case where the $T_x$ entry is empty, the column density is calculated at the
excitation temperature of the other molecule.}
\begin{tabular}{@{}l c c cc cc  @{}} 
\hline
System &  $z_{\rm abs}$ & Reference & \multicolumn{2}{c}{CO} &  \multicolumn{2}{c}{HCO$^+$} \\
       &           &              & $T_x$ [K] & $N_{\rm total}$ [\scm]& $T_x$ [K] & $N_{\rm total}$ [\scm]\\
\hline
0218$+$357 &0.68466  & \citet{wc95}  &  9 & $6.3\pm0.3\times10^{16}$  & --  &   $5.1\pm0.2\times10^{13}$  \\
1413$+$135 & 0.24671 & \citet{wc97} & --&   $1.54\pm0.04\times10^{16}$& 8  & $2.4\pm0.4\times10^{13}$  \\
1504$+$377 & 0.67335 & \citet{wc96b} & 16&  $5.9\pm1.0\times10^{16}$  &13 &   $6.5\pm0.1\times10^{13}$ \\
1830$-$211 & 0.88582 & \citet{wc96,wc98} &  -- & $\approx6\times10^{17}$ & 8 & $2.9\pm0.3\times10^{14}$\\
\hline 
\end{tabular}
\label{columns}
\end{minipage}
\end{table*} 
These are found to be close to those derived by
\citet{wc95,wc96,wc96b,wc97} and so for our targets we adopt an excitation temperature of
$T_{\rm x}=10$ K at $z=0$. %$T_{\rm CMB} = 2.73 \,(1 + z)$

\begin{figure}
\centering
\includegraphics[angle=270,scale=0.75]{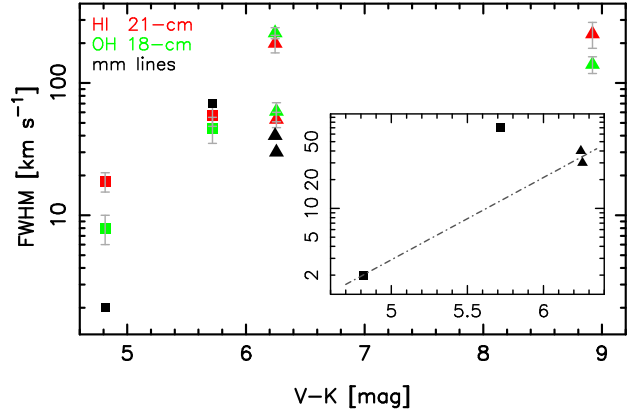}
\caption{The FWHM of the \HI\ and OH profiles versus the optical--near-infrared colour for the five
OH absorbers. The black symbols show the approximate millimetre FWHM versus
the optical--near-infrared colour, for the four of the five detected
in millimetre lines,. The inset shows the millimetre detections only with the line showing the fit used to estimate the
FWHM for our targets.}
\label{HI-OH-FWHM}
\end{figure}

Although the excitation temperatures are similar, the four known millimetre
absorbers exhibit a range of full-width half maxima (FWHMs), ranging from FWHM$\,\approx2$ \kms\ 
(1413+135, \citealt{wc97}) to $\approx70$ \kms\ (the main component towards
1504+377, \citealt{wc96b}).
For these, plus the absorber towards 0132--097 (Sect. 1), it has been shown that the OH 18-cm
($^{2}\Pi_{3/2} J = 3/2$) FWHMs are similar to those of the \HI\ 21-cm profiles \citep{cdbw07},
although we cannot yet unambiguously state that millimetre widths are correlated with those of
the decimetre lines (Fig. \ref{HI-OH-FWHM}). 
In any case, of all of our targets, only one
(0500+019)  has been detected in 21-cm absorption (see table 6 of \citealt{cww+08} and Sect. \ref{lotag}), and so we would
have no information
 on what the width of each of the undetected lines should be. If this were known, we would smooth the spectral resolution ($\Delta v$) of
the data to the FWHM of the line, thus giving the best possible optical depth limit, based upon the
detection of absorption in a single channel. 
Multiplying the
resulting optical depth, scales the velocity integrated optical depth to $\tau \times {\rm
  FWHM}/\Delta v$, giving an overall scaling of $\sqrt{{\rm FWHM}/\Delta v}$.

Choosing either the 
minimum and maximum FWHM of the four known absorbers, therefore gives a possible range in scaling factors of 
$\sqrt{70/2}\approx6$, and so the choice of either FWHM$\,\approx2$ \kms\ or $\approx70$ \kms\ could
lead to a significant over/under-estimate in the profile width. We therefore
use the optical--near-infrared colour, which appears to be correlated with the profile width in the four known systems
(Fig. \ref{HI-OH-FWHM})\footnote{This suggests that the path length through the dust responsible for the reddening 
may be correlated with the orientation of a rotating cloud complex, such
as the disk of a galaxy (Curran \& Whiting, in prep.). There
is a 1\%  probability of this correlation arising by chance (for \HI, Fig. \ref{HI-OH-FWHM}), i.e. a $3.30\sigma$
significance assuming Gaussian statistics.},  
in order to estimate what the FWHM of an undetected
absorber should be.

As seen from the figure, the associated absorber in 1504+377 represents a stray point with a FWHM of $\approx70$
\kms. We therefore estimate the profile widths from the least-squares-fit to the other three systems (shown in the inset
of Fig. \ref{HI-OH-FWHM}).  The fact that this is based upon only three points, with the removal of one which we simply
do not like, limits
the robustness of this method, although it should give a better handle on the profile width than a simple assumption.
We therefore use this fit, $\log_{10}{\rm FWHM} \approx0.86\,(V-K)-3.84$ (for $2 \leq {\rm FWHM} \leq 70$ \kms), to
estimate the column density limits (given in Table \ref{sum}).\footnote{In order to guard against over-compensating, we
  use the minimum and maximum observed profile widths of 2 and 70 \kms\ in cases where the estimated FWHMs are less or
  greater than these limiting values.} 

\subsection{Comparisons with the known high redshift molecular absorbers}
\label{comp4}

\begin{figure}
\centering
\includegraphics[angle=270,scale=0.75]{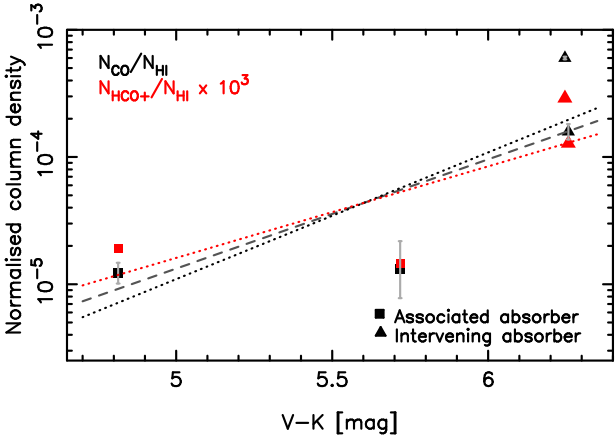}
\caption{The normalised (assuming $T_{\rm s}/f =100$ K) CO (black)
  and HCO$^+$ (coloured) column densities versus optical--near-IR
  colour. The dotted lines show the least-squares fit for each of the
molecules with the dashed line showing the composite fit.}
\label{norm-colour}
\end{figure}

\begin{figure*}
\centering
\hspace*{-2.5mm}\includegraphics[angle=270,scale=0.75]{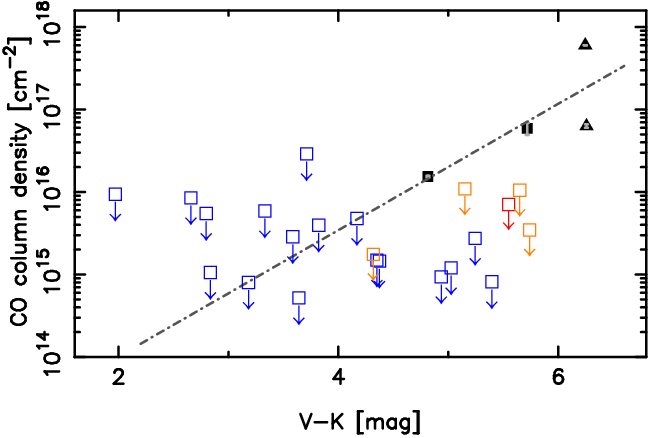}
\hspace*{+2.5mm}\includegraphics[angle=270,scale=0.75]{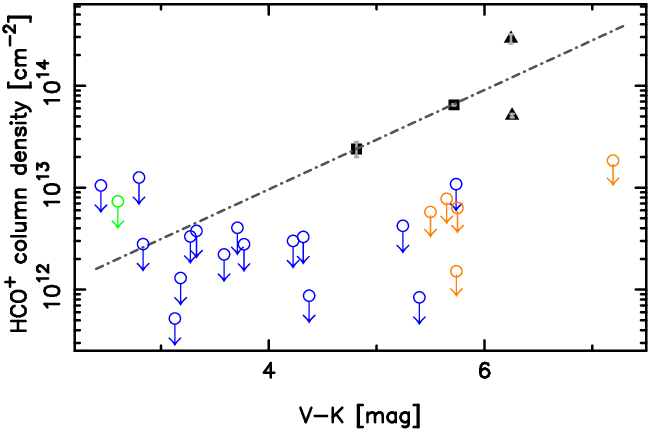}
\caption{The CO (left) and HCO$^+$ (right) column densities versus optical--near-IR colour. The
  symbols are as per Fig. \ref{norm-colour}, with the filled symbols showing the four known systems
  (the line shows the least-squares fit to these) and the unfilled symbols showing the limits from the searches by
  \citet{wc95,wc96b,dcw96} [blue] 
  \citet{mcw02} (specifically 0727--115 at $z_{\rm em} = 1.591$,
  see Sect. \ref{lotag}) [red] and the millimetre wave searches of \citet{cwm+06} [green] (actually
  HCN in 0335--122, since this is the only source for which $V$ and $K$ are available) and this paper [orange]. All of the limits are at
  the $3\sigma$ level per channel recalculated for the expected FWHM, except when outside the range of the 
four known, where either 2 or 70 \kms\ is used  (Sect. \ref{obs}). Note that this range may add a potential uncertainty
of 0.78 dex to the column density estimates (Sect. \ref{obs}, although see footnotes \ref{12} and \ref{13}), in addition to further uncertainty from the
assumption of $T_{\rm x}=10$ K at $z=0$ (see Sect. \ref{cwth}).}
\label{CO-HCO+}
\end{figure*}

In Fig. \ref{norm-colour} we show the CO and HCO$^+$ column densities
normalised by that of the \HI\ 21-cm (from \citealt{cps92,cry93,cmr+98,cdn99})
against the optical--near-IR colour, for the four known millimetre absorbers.
From this it is difficult to ascertain whether there is a correlation between the 
normalised millimetre-wave column density and the $V-K$ colour, as is
found for the normalised OH column densities in \citet{cwm+06} [Fig. \ref{frac-col}].
Furthermore, 21-cm line strengths are not available for the
sample although in \citet{cwm+06} there was still a,
albeit more scattered, correlation between the un-normalised 
OH $^{2}\Pi_{3/2} J = 3/2$ column density and the $V-K$ colour, which appears to apply to all of the
molecular absorbers (Fig. \ref{frac-col}, top).
Therefore in Fig. \ref{CO-HCO+} we show the un-normalised CO and HCO$^+$
column densities against the colour for our targets, as well as those from
previous searches of associated molecular absorption \citep{wc95,wc96b,dcw96,mcw02,cww+08}.\footnote{In the plots
  we give  \citet{wc95,wc96b,dcw96} the same colour as the latter paper includes some of the results of the first two.}
From these, we see that many of the limits (especially the new ones reported in this paper, shown in orange), should
be more than sufficient to detect molecular absorption, particularly HCO$^+$.  Incorporating the
upper limits of the column densities, via the {\sc asurv} survival analysis package \citep{ifn86}, gives Kendall's $\tau$
two-sided probabilities of $P(\tau) = 0.0092$ ($n=26$) [i.e a significance of $S(\tau) = 2.60\sigma$ for CO]\footnote{\label{12} Bearing
in mind that these limits are subject to the FWHM estimates (Sect. \ref{obs}),
we have also derived the probabilities when applying the two FWHM extrema of the detected absorbers, 2 and 70 \kms.
Not generally being as red as the detected sample, the values do not change for FWHM$=2$ \kms, and for 70 \kms, 
$P(\tau)_{70} = 0.0075$ \& $S(\tau)_{70} = 2.68\sigma$  ($n=28$).}
and $P(\tau) = 0.025$ ($n=27$) [$S(\tau) = 2.24\sigma$ for HCO$^+$]\footnote{\label{13} Again, applying
 FWHM$=2$ \kms\ introduces no change to the correlation and $P(\tau)_{70} = 0.022$ \& $S(\tau)_{70} = 2.29\sigma$  ($n=30$).}
of the column density--colour
correlations occuring by chance.  For the five known OH absorbers, $S(\tau) = 1.96\sigma$ \citep{cwm+10}, which
falls to $1.36\sigma$ for those four detected in millimetre transitions.\footnote{$S(\tau) = 2.25\sigma$
for the five known OH absorbers plus the OH limits \citep{cwm+10}.} 
So although the search
results do not lessen the significance of the column density--colour correlations (but actually
increase them), many of the limits are below those expected for a detection and so
we discuss other possible reasons as to why millimetre-wave spectral lines were undetected in this and
the previous surveys.

\section{Discussion: Possible explanations for these and the previous non-detections}
\subsection{Millimetre-wave covering factors}
\label{mwcf}

While there is an apparent correlation between the normalised column density and the optical--near
infrared colour, Fig.~\ref{frac-col} actually shows the ratio of the OH 18-cm velocity
integrated optical depth to that of the \HI\ 21-cm. Therefore in order for the fit to be accurate
for the normalised column densities, the four known millimetre absorbers must also have similar spin
and excitation temperatures, as well as \HI\ and OH covering factors (or at least a similar ratio of
these quantities).  While the excitation temperatures can be well constrained from the millimetre
transitions (Table~\ref{columns}) and the spin temperatures can be removed by not normalising the
column densities (Fig.~\ref{norm-colour} cf.~\ref{CO-HCO+}), the covering factor, $f$, is ingrained
into the optical depth via, $\tau=-\ln\left(1-\frac{\sigma}{f\,S}\right)$.  Since the CO and HCO$^+$
lines are optically thick in the four known systems, $-\ln\left(1-\frac{\sigma}{f\,S}\right)\neq
\frac{\sigma}{f\,S}$, thus not allowing $f$ to be taken outside of the integral in Equation~\ref{equ2}, 
although, as noted in Sect. \ref{obs}, if $f\approx1$ this would have little effect.

In any case, the correlation shown in Fig. \ref{norm-colour} does not hold
for {\em all five}  of the absorbers: 
Towards the $z_{\rm em} =
2.215$ quasar PMN J0134--0931 (0132--097), \citet{kcl+05} detected OH
at a column density of $N_{\rm OH}=5.6\times10^{14}.\,(T_{\rm x}/f)$
\scm\ in the intervening $z_{\rm abs} = 0.765$ gravitational lens,
although HCO$^+$ $1\rightarrow2$ was undetected to $\tau<0.07$. This
gives a column density ratio of $N_{\rm OH}\gapp\,5600 N_{\rm
  HCO^+}$.
\citet{cdbw07} have suggested that the non-detection of
millimetre absorption towards 0132--097 could be a geometrical effect,
where a molecular cloud has a much larger chance of occulting the
lower frequencies, since the decimetre emission region is likely to be much larger
than the millimetre region (as may also be exhibited by the wider decimetre profiles, Fig. \ref{HI-OH-FWHM}).
That is, the narrower millimetre emission is less likely to intercept the CO and HCO$^+$ gas which
will be localised in dense molecular cores. The fact that the millimetre and decimetre emission, and thus absorption, trace
different components of the gas could be responsible for the significance of the correlation with colour 
not changing when normalising the millimetre column densities by
that of the \HI\ (Sect.~\ref{comp4}).

In the case of 0132--097 , the possibility that the core of the
absorbing molecular gas is offset from the centre of the emission may
also result in a reduction of the decimetre covering factor, which
could be the cause of the normalised OH line strength in 0132--097 not
being as high as expected from its colour (i.e. below the fit defined
by the others in Fig.~\ref{frac-col}, right), although more absorbers
would be required to verify this.

\subsection{Location of the absorbing gas}
\label{lotag}

 From the fact that, of our sample, \HI\ has only been detected in 0500+019\footnote{As well as a possible
detection towards 1107--187, which at $z=0.48909$ \citep{cwm+10}, is not covered by the 76.88--77.12 GHz
($0.4947 < z_{\rm CO} < 0.4994$) range observed here (Fig. \ref{spectra}). 0213--026, 1504--166 and 1706+006 
are undetected in 21-cm \citep{cwm+10}, with the remainder of the sample as yet unsearched (see tables 1 \& 2 of \citealt{cw10}). }
we may be incorrect in our assumption that the host
galaxy is generally the location of the quasar reddening, as it is for only
two of the five known OH absorbers. As is the case for the remaining three of the
known systems, the absorption (and thus the cause of the reddening) could be
occuring anywhere between us and the emission redshift of the
quasar. Unfortunately, full range spectral scans are not feasible
towards most sources with current instruments and,
over the ranges which are scanned, there is a trade-off between
the redshift space which can be covered and the depth of the
search. Thus our motivation for searching for molecular absorption
within the host galaxy. Spectral scans have, however, been attempted
\citep{mcw02,cwmk03}, with 0500+019 being one of the targets common
to ours. We summarise
the results in Table \ref{scans},
\begin{table*} 
\centering
\begin{minipage}{173mm}
\caption{Summary of the millimetre-band scans towards 
0500+019 with the Swedish ESO-Submillimetre Telescope \citep{mcw02}
in order of increasing frequency bands, as tabulated in \citet{cwmk03}. 
We also show the results for 0727---155 since the emission redshift,
 $z_{\rm em}$, of the quasar is covered (by CO $1\rightarrow2$, where we show
the fractional redshift range based upon $z =1.547 - 1.591$). This is therefore included in
  our analysis (Fig.~\ref{CO-HCO+}). 
The column density limits are calculated from
the quoted optical depths per a 18 \kms\ (0500+019) and 70 \kms\ (0727--115) channel (Sect.~\ref{obs}), with $T_{\rm x} = 10$~K at $z=0$.}
\begin{tabular}{@{}l c c c c c c c  c cc @{}} 
\hline
Quasar & $V$ & $z_{\rm em}$ & &\multicolumn{4}{c}{HCO$^+$} & \multicolumn{3}{c}{CO} \\
 &     &         &     Trans. & $z$-range & $\Delta z/z_{\rm em}$ & $N_{\rm total}$ [\scm]& Trans. & $z$-range & $\Delta z/z_{\rm em}$ & $N_{\rm total}$ [\scm]\\
\hline
0500+019 & 21.2 & 0.58457 & $0\rightarrow1$ & 0.1025--0.1391 & 0.063 & $<4.1\times10^{13}$& $0\rightarrow1$ &0.4249--0.4722 & 0.081 & $<3.0\times10^{16}$\\ %78.30000 to 80.90000 GHz
... & ...& ...&$0\rightarrow1$  & 0.0311--0.0431 & 0.021 & $<2.3\times10^{13}$ & $0\rightarrow1$ &0.3326--0.3482 & 0.027 & $<1.7\times10^{16}$\\%85.50000 to 86.50000 GHz
... & ...& ...& $1\rightarrow2$ &0.5771--0.5846  &  0.013 & $<1.0\times10^{13}$& $0\rightarrow1$  & 0.0192--0.0283 & 0.016 & $<1.7\times10^{16}$ \\ %112.10000 to 113.10000 GHz
... & ...& ...& $1\rightarrow2$ & 0.2615--0.3721 &  0.189 &$<1.2\times10^{13}$ &  --- & ---& --- & --- \\%130.00000 to 141.39999 GHz
%%%%%%%%%%%%%%%%%%%%%%%%%%%%%
0727--115 & 22.5 & 1.591 & $0\rightarrow1$ &0.1025--0.1391 & 0.023 & $<4.3\times10^{13}$& $0\rightarrow1$ &0.4249--0.4722 & 0.030 & $<3.2\times10^{16}$\\
... & ...& ...& $1\rightarrow2$ &1.2049--1.2781 &  0.046 & $<1.7\times10^{13}$ &  --- & ---& --- & --- \\ %78.30000 to 80.90000

... & ...& ...&   $0\rightarrow1$ & 0--0.0630 & 0.040 &  $<2.1\times10^{13}$& $0\rightarrow1$ & 0.2737--0.3739 & 0.063 & $<1.6\times10^{16}$\\
... & ...& ...&    $1\rightarrow2$ &0.9710--1.1260 & 0.097 &$<8.5\times10^{12}$ & $1\rightarrow2$ & 1.5474--1.7478& 0.027 & $<7.1\times10^{15}$ \\ %83.90000 to 90.50000 GHz

... & ...& ...&  $1\rightarrow2$ &0.2615--0.2926 & 0.020 &$<1.7\times10^{13}$ & $1\rightarrow2$ & 0.6304--0.6706 & 0.025 & $<1.5\times10^{16}$\\
... & ...& ...&$2\rightarrow3$ &  0.8922--0.9388 &  0.029 &$<1.7\times10^{13}$ & $2\rightarrow3$ &1.4455--1.5058 & 0.038& $<1.8\times10^{16}$ \\
... & ...& ...&$3\rightarrow4$ & 1.5229--1.5850 & 0.039 &$<2.9\times10^{13}$ & --- & ---& --- & --- \\%138.00000 to 141.39999 GH

... & ...& ...&  $1\rightarrow2$ & 0.2004--0.2491 & 0.031 & $<1.5\times10^{13}$& $1\rightarrow2$ & 0.5514--0.6144 & 0.040 & $<1.3\times10^{16}$ \\
... & ...& ...&$2\rightarrow3$ & 0.8005--0.8737 & 0.046 &  $<1.4\times10^{13}$& $2\rightarrow3$ &1.3270--1.4215 & 0.059 & $<1.5\times10^{16}$ \\
... & ...& ...&$3\rightarrow4$ &1.4006--1.4981 & 0.061 & $<2.5\times10^{13}$& --- & ---& --- & --- \\%142.80000 to 148.60001 GHz
\hline
\end{tabular}
\label{scans}
\end{minipage}
\end{table*} 
from which we see that $<30\%$ of the redshift space towards 0500+019 has been scanned for HCO$^+$ and only 12\% for CO, with some overlap in these
ranges. We therefore cannot rule out that the cause of the reddening may be at some other redshift towards 0500+019 and indeed for the rest of the
sample.

There is also the possibility that, even if the reddening (and hence any absorption) is located at the host, it would only be detected if the optical
redshift is known precisely enough to ensure that the observed band covers the correct redshift range.
The optical spectra of these quasars have spectral resolutions of $\approx5-8$~\AA, which corresponds to a few hundred \kms,
which is close to our observed bandwidths (Fig. \ref{spectra}). However, most of the quasars observed here
have have strong narrow emission lines, which constrain the redshift to better than $\sim10^{-4}$ 
and all are quoted to at least the third decimal place, although uncertainties are usually not given \citep{dwf+97}.
Referring to the ranges covered by our observed bands (Table \ref{sum}), we see that
offsets of $\Delta z \approx\pm 0.002$ (at low redshifts) to $\approx\pm0.01$ (at high redshifts) are covered,
thus making it unlikely that our observations are generally tuned to the wrong frequency.

%\newpage
\subsection{Conditions within the host}
\label{cwth}

Above we suggest the possibility that molecular absorption is not detected since the reddening of the quasar light may
not necessarily be occuring close to the redshift of the background quasar.  Over and above this, there is the
possibility that a large column of absorbing molecular gas {\em cannot} be located in the host galaxy on the basis that
the there is a low fraction of cool neutral gas in the hosts of quasars and radio galaxies with ultra-violet
luminosities of $L_{\rm UV}\gapp10^{23}$ \WpHz\ (\citealt{cww+08}). Whether this is due to excitation of the gas or
these luminosities selecting gas-poor ellipticals (or indeed whether these two possibilities are inter-related,
\citealt{cw10}), this appears to be the case for {\em all} redshifted ($z_{\rm em} \gapp0.1$) \HI\ 21-cm absorption
searches. % \citep{cwm+10}.

\begin{figure*}
\centering \includegraphics[angle=270,scale=0.75]{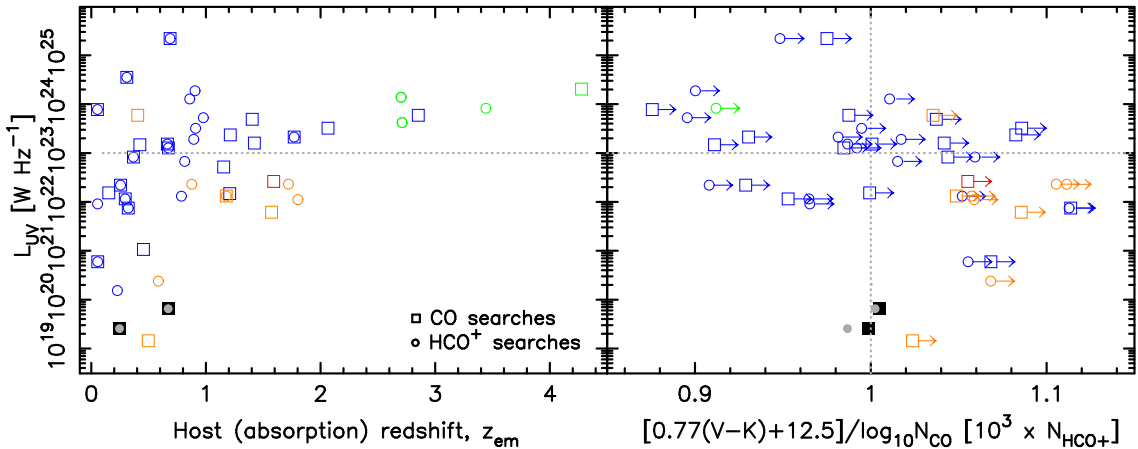}
\caption{The ultra-violet ($\lambda\approx1216$ \AA) luminosity versus the redshift [left] and 
``{\em colour depth}'' [right] for the sources listed in Table \ref{all-mags}. The {\em colour
  depth} is defined by the fit to the CO column density--colour correlation for the four known systems (Fig. \ref{CO-HCO+}, left), i.e.
$[0.77\,(V-K) + 12.5]/\log_{10}N_{\rm CO}$ and $0.77\,(V-K) + 12.5/\log_{10}(1000 \times N_{\rm
  HCO^+})$, with the two known associated systems defining the vertical line at a value of 1.0. 
In the left panel we show all of the sources searched, as $V$ and $K$ magnitudes
are not required, nor is the re-calculated $3\sigma$ r.m.s. noise required to be lower than
the flux (as is not the case for the PKS 0422+004 searches nor HCO$^+$ in Hydra A, \citealt{dcw96}).
Note that, after the $\times1000$ normalisation, the  HCO$^+$ points (shown in grey for clarity) are located close to the CO
points for the two known associated systems. }
\label{lum}
\end{figure*}

\begin{table*} 
\centering
\begin{minipage}{154mm}
\caption{The magnitudes and calculated $\lambda\approx1216$ \AA\ luminosities [\WpHz] of the 
quasars and radio galaxies searched in millimetre-wave absorption. The bottom panel lists the two known associated absorbers. }
\begin{tabular}{@{}l l c c c c c c c c @{}} 
\hline
Quasar &  $z_{\rm em}$ & Molecules                            & Ref &  $B$ & $V$ &$K$ & Refs  & $\log_{10}L_{\rm UV}$\\
\hline
4C\,+40.01        & 0.255   & CO, HCO$^+$                 & W96 &   17.89  &    17.67 &  14.87  & Z04,S06  &  22.34   \\ % B3 0010+405
PKS 0113--118  &  0.672  & CO, HCO$^+$                 & D96 &   18.27  &  17.95  &  14.37  &  Z04     &     23.11 \\ % Lum includes R=17.72 (Z04) in W96 too 			  
PKS 0213--026  &  1.178 & CO, HCO$^+$                & C11 &   21.33   &  20.82  &  15.17   &  F00      &    22.12\\% R = 20.475 GBT paper -HO1			  
\relax[HB89] 0234+285 & 1.213 &  CO                      &W95 &   19.20 &  18.35 &  13.33 & A09, S06   &   23.37 \\         
PKS 0335--122 &  3.442 & HCN                                  & D96 &  21.02 & 20.11 & 17.51  & E05,H01  &    23.92 \\							   
\relax[HB89] 0422+004  & 0.310  & CO, HCO$^+$, HCN        & D96 &  13.89   & 14.12  & 11.31  & F04        &  24.55  \\ %Optically variable - B&V averages		  
PKS 0434--188 & 2.702 & HCO$^+$, HCN                & C08 &  19.25    &  ---     &  16.24   & E05         &  24.14 \\							  
PKS 0438--436  &  2.852 & CO                                   & D96 &  20.74   & 19.91   & 16.09  &  B04,E05     &  23.77\\							  
PKS 0446+11     & 1.207   & CO                                   & W96 & 21.43&  20.36  & 15.37 &  A09,S06  &    22.17 \\           
PKS 0454+06  &  0.405  & CO                                     & C11 &  19.38   &  18.74 & 14.42  &   F00       &    21.57 \\%R = 18.589 GBT paper -HO1			   
PKS 0500+019  &  0.58457& HCO$^+$                     & C11 &  22.50  & 21.35   & 15.6   &  D97,C03,S96b  & 20.38  \\ %R =20.682 - C03				  
PKS 0521--365  &  0.0552 & HCO$^+$                     & D96 &  15.60   & 14.60  &  11.33  & H07,S06     &  21.96 \\							  
PKS 0528+134   & 2.065     &  CO                              &W95 &     20.09  &  20.00  & 15.06 & Z04,G00    &  23.51  \\
\relax[HB89]  0537--441  &  0.894  & CO, HCO$^+$,HCN,CS & D96 &  17.93   & 17.34   & 13.02   & F00         &  23.28\\							  
PKS 0601--17 & 2.711 & HCO$^+$, HCN                  & C08 &  20.45  &  ---    &   ---     & H01     &      23.66 \\							  
CGRaBS\,J0650+6001 & 0.455 & CO                            & W96 &   20.85 &   ---     & 14.63  & Z04,S06    &  21.03 \\
PKS 0727--115  & 1.591  & CO                                   & M03 & 21.13 & 20.07  & 14.52  & A09,C03  &     22.42 \\% B & V converted from SDSS ugriz. 		  
\relax[HB89] 0735+178   & 0.424   &  CO                  &  W95 &  16.30 &  15.68 &  13.03  & A09,S06  &    23.17 \\
PKS 0823-223        & 0.9103 & HCO$^+$                 & W96 &     ---    & 16.11  & 12.34 &  F93   &       23.51\\  
Hydra A	  & 0.0538  & CO, HCO$^+$                         & D96 & 13.90   & 12.87   & 10.90  &  V91,C03    &   23.89 \\ % 09h18m05.7s					   
\relax[HB89]  0954+658  & 0.368&  CO, HCO$^+$       & W96 &  17.18  & 16.80  & 12.43  & P04,S06 &     22.92 \\    
PKS 1026--084 &  4.276 & CO                                    & C08 & 21.07  & ---    &   ---      & H01      &     24.31 \\%R = 19.154  -H01					  
PKS 1107--187  &   0.497 & CO                                  & C11 &  22.44  &  21.10 &   15.95   & F00      &     19.16\\%R = 19.509- H01					  
FTM\,J1341+3301 &   1.720 & HCO$^+$                   & C11 &   23.39  & 21.68  & 14.49  & A09,G07     &  22.36 \\ % B & V converted from SDSS ugriz. Lum from SDSS.	  
SBS 1347+539B     & 0.978   & HCO$^+$                 & W96 &  17.95  & 17.52  & 15.07  & A09,S06  &    23.72 \\          
\relax[HB89] 1418+546 & 0.152  &  CO                   & W96 &     16.25 & 15.60&  11.43 & A09,S06   &   22.18 \\ %warning - variable!   
PKS 1430--155  &   1.573 & CO                                  & C11 &  22.50  & 23.24  & 17.50 &  D97,F00  &     21.79 \\ %R = 22.910 - F00 				  
3C\,309.1            &    0.905  &  CO                             & W96 &   16.80   &  16.57    & 14.72  &  Z04,S06   &   24.27                  \\ % [HB89] 1458+718
PKS 1504--167   &  0.876 & HCO$^+$                      & C11 &  20.28   & 19.75  & 14.01  & F00       &    22.36 \\%R=19.350 -   F00 				  
PKS 1548+056  &  1.422  &  CO                                  & D96 &  19.37   & 18.65   & 14.30   & F00          & 23.21\\							  
PKS 1555+001  &  1.77  &  CO, HCO$^+$                 & D96 &  20.34   & 19.95   & 16.24   & F00          & 23.33\\							  
PKS 1622--253 & 0.786   & HCO$^+$                        & W96 &      ---   & 20.6  &  14.86 & d94,S06   &   22.12 \\ %R=20.3  
PKS 1622--29   & 0.815  &  HCO$^+$                       & W96 &     ---    &18.38   &14.15  &R02,S06    &  22.83 \\
PKS 1725+044  &  0.296 &  CO, HCO$^+$                & D96 &  17.92   & 17.46  &  14.13   & F00         &  22.06\\							  
\relax[HB89] 1749+096 & 0.322 & CO, HCO$^+$     & W96 &   17.98 &  17.39 &  12.00&   O09,S06    &  21.87\\        
\relax[HB89] 1823+568 & 0.664 & CO, HCO$^+$    & W96 &    17.41 &  17.02  & 13.84 &  Z04,S06    &  23.18 \\    
Cygnus A   &  0.0561& CO, HCO$^+$                        & D96 & 17.04   & 15.52   & 10.28 & V91,C03     &  20.78\\	%1957+406						   
CGRaB\, J2022+6136 & 0.228 & HCO$^+$                 & W96 &    19.55   &17.97  & 14.28 & Z04,S06    &  20.19 \\ %87GB[BWE91] 2021+6127      
\relax[HB89] 2200+420   & 0.688  & CO, HCO$^+$   & W96 &    13.25&   13.32  & 10.49 & F04,S06    &  25.34 \\ % BL Lac itself!     
PKS 2223--052 &  1.404 & CO                                   & D96  & 18.59   & 18.33   & 14.69   & F00       & 23.69\\							   
3C\,454.3           & 0.859 &     HCO$^+$                      &W95 &     16.03 &  16.19  & 13.06 & Z04,S06     & 24.11\\ % [HB89] 2251+158   
PKS 2329--162 &  1.155 & CO                                   & D96  & 20.72  & ---      & 16.58   & H01,W83    &   22.72 \\							   
PKS 2337--334   &  1.802 & HCO$^+$                      & C11  & 22.93   & 21.89   & 16.39 &   F00     &      22.05 \\							  
\hline                                                          
PKS 1413+135  &  0.246710 & CO, HCO$^+$, HCN, HNC & W97& 21.37&  19.74 &  14.93 &  A09,S06  &     19.41 \\  % lum from SDSS: ugriz=21.94 20.53 19.05 18.50 18.09 
B3 1504+377  &  0.67150  & CO, HCO$^+$, HCN, HNC & W96 & 23.24 &  21.82  & 16.10  &  A09,S96a     &  19.81 \\  % lum from SDSS
\hline       
\end{tabular}
{Molecular search references: D96 -- \citet{dcw96}, W95 -- \citet{wc95},  W96 -- \citet{wc96b}, W97 -- \citet{wc97}, M03 --
  \citet{mcw02},  C08 -- \citet{cww+08}, C11 -- this paper.\\
Photometry references: W83 -- \citet{waa83}, F93 -- \citet{fbbt93}, d94 -- \citet{ddmt94}, S96a -- \citet{srkr96}, S96b -- \citet{srrk96},  D97 -- \citet{dwf+97},  F00 -- \citet{fww00}, G00 -- \citet{grss00}, E01 -- \citet{eyh+01}, H01 --  SuperCOSMOS Sky Survey \citep{hmr+01},  R02 -- \citet{rcca02}, C03 -- \citet{cb03}, B04 -- \citet{bpm+04}, F04 -- \citet{fct04}, P04 -- \citet{psbp04}, Z04 -- \citet{zml+04}, E05 -- \citet{ehl05}, S06 -- 2MASS \citep{scs+06}, A09 -- SDDS DR7\citep{aaa+09},  O09 -- \citet{ozh+09}, FPC -- P.~Francis (priv. comm.).}
\label{all-mags}  
\end{minipage}
\end{table*} 
In order to investigate whether this could affect our sample, we use the photometry
of the searched sources (Table \ref{all-mags}) 
 to derive the UV luminosities (as per \citealt{cww+08}). Showing these in Fig.~\ref{lum},
we see the expected increase in luminosity with redshift and note that our selection of the reddest
PHFS sources (Sect.\ref{ts}) means that nearly all of our targets lie below $L_{\rm
  UV} = 10^{23}$ \WpHz \ (left panel).
Since we are unlikely to detect molecular absorption
in an environment not conducive to large columns of cool neutral gas,
 in the right panel of Fig. \ref{lum} we show the  UV luminosity versus the {\em colour depth}, which we define as the
optical--near-infrared colour normalised by the depth of the search. As expected from the column
density---$V-K$ correlation (Fig. \ref{norm-colour}), this gives a vertical line with
this quantity plotted as the abscissa\footnote{Adding the three intervening absorbers gives a {\em
    colour depth} of $1.00\pm0.02$ for  both $N_{\rm CO}$ and $1000 \times N_{\rm HCO^+}$.
    The intervening absorbers  are not shown in the plot as, although the ultra-violet luminosities can also be
    estimated, these are remote from the source.}
  and from this definition we see that 20 discrete sources
(mostly from this  work) have been searched  sufficiently deeply, of
which only 12 are located 
in the bottom right quadrant defined by $L_{\rm
  UV}\lapp10^{23}$ \WpHz\ and a {\em colour depth }$\gapp1$.

As expected, in Fig.~\ref{lum} (right) there is an anti-correlation between the UV luminosity and $V-K$ colour, since
the former is derived from the observed frame optical photometry. As such, there is also the possibility that the
calculated values of $L_{\rm UV}$ are influenced by the dust extinction
towards the source. However, the fact that 21-cm absorption is never detected above a critical luminosity (or below a
critical magnitude/dust extinction), as well as the correlation between the 21-cm absorption strength and $V-K$ colour
\citep{cw10}, indicates that the extinction occurs within the host in the case of the 21-cm detections, which follow the
expected 50\% detection rate for $L_{\rm UV}\lapp10^{23}$ \WpHz\ \citep{cw10}.  Therefore, whether due to high intrinsic
UV luminosities or a paucity of dust within the host galaxy, it remains that 21-cm absorption is not detected where
$L_{\rm UV}\gapp10^{23}$ \WpHz\ and where we do not detect 21-cm we do not expect to detect absorption by molecular gas.

\begin{figure}
\centering \includegraphics[angle=270,scale=0.45]{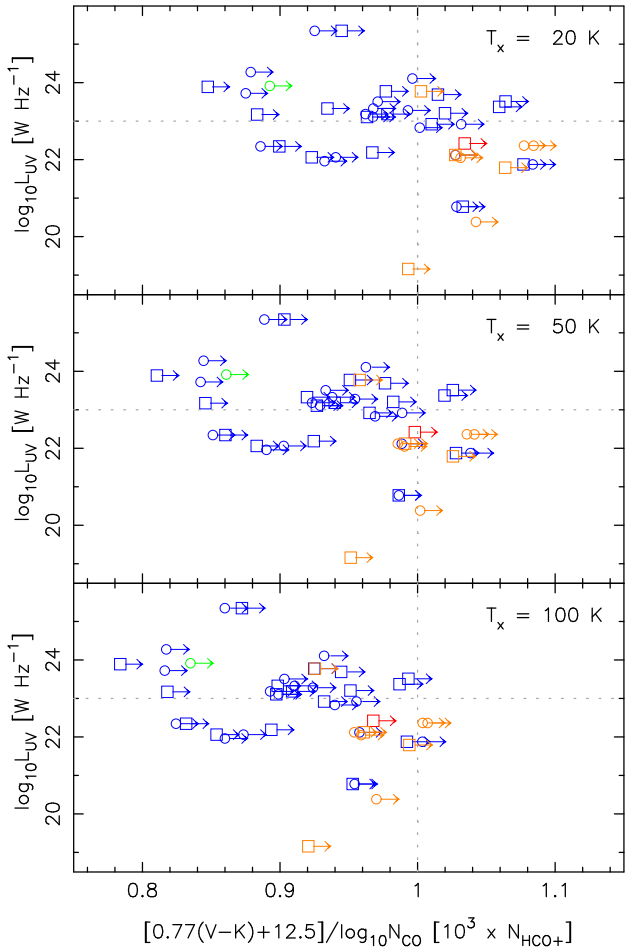}
\caption{The {\em colour depth} distribution for various excitation temperatures.}
\label{3-lum}
\end{figure}

Lastly, as well as a covering factor of $f\approx1$, as per the four known systems, we have assumed an
excitation temperature of $T_{\rm x}\approx 10$ K (at $z=0$) [Sect. \ref{mwcf}]. Since there is no
apparent correlation between the optical--near-infrared colour and the excitation
temperature for the four known systems (Table \ref{columns})\footnote{For 1413$+$135, $V-K = 4.81$ and $T_{\rm x}=8$ K,
similar to the excitation temperature of  1830$-$211 with $V-K = 6.25$.}, we cannot estimate
temperatures for any of the searched sample. If the gas were at higher excitation temperatures, being
mostly observations of the lower rotational transitions (Table \ref{sum}), the limits are less
sensitive than quoted since a larger total column
density would be required to give the same observed flux in these transitions. This has the effect of worsening the limits to
the point where only two of the searched sources are (just) potentially detectable when $T_{\rm x}\approx 100$~K (Fig. \ref{3-lum}),
although if the gas is diffuse, such excitation temperatures may not be attainable, even in the presence of high kinetic temperatures.

\section{Summary}

We have undertaken a survey for molecular absorption in the millimetre-band at high redshift. Like
all previous surveys (since \citealt{wc95,wc96,wc96b,wc97}), we have not detected millimetre-band
spectral line absorption in any of the targets. In the case of absorption due to intervening
sources, it has been established that the paucity of detections is due to the traditional optical
selection of these objects, in which the redshift of the intervening object is usually known,
biasing against the reddest, and thus dustiest, absorption complexes (\citealt{cwm+06}). We have
therefore aimed to circumvent this bias, by searching for molecular absorption within the hosts of
particularly red objects, although like \citet{dcw96}, who also targetted red sources (and type-2
AGN)\footnote{Although as \citet{cw10} have shown, the bulk cool neutral gas (as traced by \HI\ 21-cm
  absorption) is not located in the obscuring torus, invoked by unified schemes of active galactic
  nuclei.}, we find no absorption. 

A possible reason for some of the
non-detections is that only 12 of the 40 objects searched are known to be located in the bottom right
quadrant of the ultra-violet luminosity--{\em colour depth} plot (Fig. \ref{lum}, right), i.e. where
$L_{\rm UV} \lapp10^{23}$ \WpHz\ and $[0.77\,(V-K) + 12.5]/\log_{10}N_{\rm CO} \approx [0.77\,(V-K)
+ 12.5]/\log_{10}(1000 \times N_{\rm HCO+})\gapp1$.  This suggests that only a dozen of the sources may be
subject to an ultra-violet flux which permits the presence of cool neutral gas \citep{cww+08}, while
being ``sufficiently red'' to indicate the presence of a large column of dust along the sight-line
to the quasar \citep{cwm+06}.  Given that:
  %    \begin{enumerate}
\begin{itemize}
       \item[--] For these, the reddening could be occuring anywhere along the 
   line-of-sight, and not necessarily in the host galaxy, as is the case for three of the five known OH
 absorbers (four of which make up all the known redshifted millimetre-band absorbers).
  The only reason we chose to observe the emission redshift is
   that it gives a frequency to which to tune the receiver --
   \citet{mcw02,cwmk03} previously attempted millimetre-wave spectral scans
   towards very dim objects, but the redshift range scanned is a trade-off with the 
   time spent on each frequency, resulting in poor optical
   depth limits over what were limited redshift ranges in only a few sources. 

\item[--]  This fifth absorber, which was detected in OH but not HCO$^+$, may be the result of
         lower millimetre-wave covering factors in comparison to those
         at decimetre wavelengths \citep{cdbw07}. 
This may be analogous to the effect seen in the optically selected DLAs
(Sect. \ref{sec:intro}), where, due to the steeper cosmological evolution
in the heavy element abundance in the \MOLH-bearing DLAs, 
\citet{cwmc03} suggest that these
constitute a more homogeneous class of objects than the general
population DLA \citep{pgw+03}.  That is, \MOLH\ absorption may only be observed
in a limited subset of possible sight-lines and \citet{zp06} suggest
that a distinction could arise from the much smaller cross-section of
the molecular gas, located in small and dense regions (e.g. \citealt{ll96a}), in
comparison to that of the atomic gas. This means that only a narrow
sight-line will occult the quasar, although the DLA may be apparent
through the more widespread 21-cm absorption.

\item[--] As well as covering factor effects, if the excitation temperature of the putative
  absorbing gas is higher than the assumed $T_{\rm x} = 10$ K (at $z=0$), the column density limits
  are poorer than those calculated here. If large enough ($T_{\rm x} \gapp100$ K), none of these
  searches, mostly of the lower rotational transitions, have been sensitive enough to detect CO or
  HCO$^+$ absorption in these sources, although the absorbing medium may be too diffuse
 to reach such temperatures through collisional excitation.

%\end{enumerate}
\end{itemize}

%\end{itemize}
\citet{dcw96} accounted for their non-detections by suggesting that the extinction may be occuring outwith the host galaxy or that the X-ray flux from
the AGN may be photo-dissociating the molecules. These possibilities are similar to the points made above, although by quantifying these via the
findings of \citet{cwm+06,cww+08}, we find that not all of the targets of \citet{dcw96} are sufficiently faint and reddened to have been detected, no
matter the location of the obscuring material.  Therefore, the key to finding new redshifted molecular absorbers in the  millimetre-band is through the  selection
of the faintest optical objects (giving $L_{\rm UV} \lapp10^{23}$ \WpHz) and performing spectral scans towards these to {\em colour depth} limits of
$\gapp1$, as defined by the column densities and colours of the four known absorbers. In order to also circumvent the covering factor effect, pilot
searches for OH in the decimetre band could be undertaken, with the low rest frequency (1667 MHz) allowing a full spectral scan in only four separate
tunings with the Square Kilometre Array \citep{cdk04}.

\section*{Acknowledgements}
We would like to thank A. Weiss for assisting with the IRAM observations, as well as Shigeru Takahashi, Jun Maekawa and Hiroki Ashizawa of Nobeyama
Radio Observatory for their assistance in installing {\sc newstar} and the data retrieval.  
MTM thanks the STFC for an Advanced Fellowship and the
Australian Research Council for a QEII Research Fellowship (DP0877998).

We acknowledge financial support from the Access to Major Research
Facilities Programme which is a component of the International Science
Linkages Programme established under the Australian Government's
innovation statement, Backing Australia's Ability.

This research has made use of the NASA/IPAC Extragalactic Database (NED) which is operated by the Jet Propulsion Laboratory, California
Institute of Technology, under contract with the National Aeronautics and Space Administration. This research has also made use of NASA's
Astrophysics Data System Bibliographic Services and {\sc asurv} Rev 1.2 \citep{lif92a}, which implements the methods presented in
\citet{ifn86}.

Funding for the Sloan Digital Sky Survey (SDSS) and SDSS-II has been
provided by the Alfred P. Sloan Foundation, the Participating
Institutions, the National Science Foundation, the U.S. Department of
Energy, the National Aeronautics and Space Administration, the
Japanese Monbukagakusho, and the Max Planck Society, and the Higher
Education Funding Council for England. The SDSS Web site is
http://www.sdss.org/.

%\bibliographystyle{../15/apj} % start with this as mn2e does not distinguish a,b, etc.
%\bibliographystyle{mn2e}
%\bibliography{aa,ref}

\begin{thebibliography}{}

\bibitem[\protect\citeauthoryear{{Abazajian}, {Adelman-McCarthy},
  {Ag{\"u}eros}, {Allam}, {Allende Prieto}, {An}, {Anderson}, {Anderson},
  {Annis}, {Bahcall} \& et al.}{{Abazajian} et~al.}{2009}]{aaa+09}
{Abazajian} K.~N.,  {Adelman-McCarthy} J.~K.,  {Ag{\"u}eros} M.~A.,  et al., 2009, ApJS, 182, 543

\bibitem[\protect\citeauthoryear{Becker, White \& Helfand}{Becker
  et~al.}{1995}]{bwh95}
Becker R.~H.,  White R.~L.,    Helfand D.~J.,  1995, ApJ, 450, 559

\bibitem[\protect\citeauthoryear{{Brocksopp}, {Puchnarewicz}, {Mason},
  {C{\'o}rdova} \& {Priedhorsky}}{{Brocksopp} et~al.}{2004}]{bpm+04}
{Brocksopp} C.,  {Puchnarewicz} E.~M.,  {Mason} K.~O.,  {C{\'o}rdova} F.~A.,
  {Priedhorsky} W.~C.,  2004, MNRAS, 349, 687

\bibitem[\protect\citeauthoryear{{Carilli}, {Menten}, {Reid}, {Rupen} \&
  {Yun}}{{Carilli} et~al.}{1998}]{cmr+98}
{Carilli} C.~L.,  {Menten} K.~M.,  {Reid} M.~J.,  {Rupen} M.~P.,    {Yun}
  M.~S.,  1998, ApJ, 494, 175

\bibitem[\protect\citeauthoryear{{Carilli}, {Perlman} \& {Stocke}}{{Carilli}
  et~al.}{1992}]{cps92}
{Carilli} C.~L.,  {Perlman} E.~S.,    {Stocke} J.~T.,  1992, ApJ, 400, L13

\bibitem[\protect\citeauthoryear{{Carilli}, {Rupen} \& {Yanny}}{{Carilli}
  et~al.}{1993}]{cry93}
{Carilli} C.~L.,  {Rupen} M.~P.,    {Yanny} B.,  1993, ApJ, 412, L59

\bibitem[\protect\citeauthoryear{Chandra, Kegel, Roy \& Hertenstein}{Chandra
  et~al.}{1995}]{cklh95}
Chandra S.,  Kegel W.~H.,  Roy R. J.~L.,    Hertenstein T.,  1995, A\&AS, 114,
  175

\bibitem[\protect\citeauthoryear{Chandra, Maheshwari \& Sharma}{Chandra
  et~al.}{1996}]{cms96}
Chandra S.,  Maheshwari V.~U.,    Sharma A.~K.,  1996, A\&AS, 117, 557

\bibitem[\protect\citeauthoryear{{Chengalur}, {de Bruyn} \&
  {Narasimha}}{{Chengalur} et~al.}{1999}]{cdn99}
{Chengalur} J.~N.,  {de Bruyn} A.~G.,    {Narasimha} D.,  1999, A\&A, 343, L79

\bibitem[\protect\citeauthoryear{{Cody} \& {Braun}}{{Cody} \&
  {Braun}}{2003}]{cb03}
{Cody} A.~M.,  {Braun} R.,  2003, A\&A, 400, 871

\bibitem[\protect\citeauthoryear{Curran, Darling, Bolatto, Whiting, Bignell \&
  Webb}{Curran et~al.}{2007}]{cdbw07}
Curran S.~J.,  Darling J.~K.,  Bolatto A.~D.,  Whiting M.~T.,  Bignell C.,
  Webb J.~K.,  2007, MNRAS, 382, L11

\bibitem[\protect\citeauthoryear{Curran, Kanekar \& Darling}{Curran
  et~al.}{2004a}]{cdk04}
Curran S.~J.,  Kanekar N.,    Darling J.~K.,  2004a, Science with the Square
  Kilometer Array, New Astronomy Reviews 48.
Elsevier, Amsterdam, pp 1095--1105

\bibitem[\protect\citeauthoryear{Curran, Murphy, Pihlstr\"{o}m, Webb, Bolatto
  \& Bower}{Curran et~al.}{2004b}]{cmpw03}
Curran S.~J.,  Murphy M.~T.,  Pihlstr\"{o}m Y.~M.,  Webb J.~K.,  Bolatto A.~D.,
     Bower G.~C.,  2004b, MNRAS, 352, 563

\bibitem[\protect\citeauthoryear{Curran, Webb, Murphy \& Carswell}{Curran
  et~al.}{2004}]{cwmc03}
Curran S.~J.,  Webb J.~K.,  Murphy M.~T.,    Carswell R.~F.,  2004, MNRAS, 351,
  L24

\bibitem[\protect\citeauthoryear{Curran, Webb, Murphy \& Kuno}{Curran
  et~al.}{2005}]{cwmk03}
Curran S.~J.,  Webb J.~K.,  Murphy M.~T.,    Kuno N.,  2005, in Engvold. O.,
  ed., Highlights of Astronomy, Vol. 13, as presented at the XXVth General
  Assembly of the IAU - 2003 Deep searches for high redshift molecular
  absorption.
ASP Conf. Ser., San Francisco, pp 845 -- 847

\bibitem[\protect\citeauthoryear{Curran, Whiting, Murphy, Webb, Bignell,
  Polatidis, Wiklind, Francis \& Langston}{Curran et~al.}{2011a}]{cwm+10}
Curran S.~J.,  Whiting M.,  Murphy M.~T.,  et al.,  2011a, MNRAS, 413, 1165

\bibitem[\protect\citeauthoryear{Curran, Whiting, Murphy, Webb, Longmore,
  Pihlstr\"{o}m, Athreya \& Blake}{Curran et~al.}{2006}]{cwm+06}
Curran S.~J.,  Whiting M.,  Murphy M.~T.,  Webb J.~K.,  Longmore S.~N.,
  Pihlstr\"{o}m Y.~M.,  Athreya R.,    Blake C.,  2006, MNRAS, 371, 431

\bibitem[\protect\citeauthoryear{Curran \& Whiting}{Curran \&
  Whiting}{2010}]{cw10}
Curran S.~J.,  Whiting M.~T.,  2010, ApJ, 712, 303

\bibitem[\protect\citeauthoryear{Curran, Whiting, Tanna, Bignell \&
  Webb}{Curran et~al.}{2011b}]{cwt+11}
Curran S.~J.,  Whiting M.~T.,  Tanna A.,  Bignell C.,    Webb J.~K.,  2011b,
  MNRAS, 413, L86

\bibitem[\protect\citeauthoryear{Curran, Whiting, Wiklind, Webb, Murphy \&
  Purcell}{Curran et~al.}{2008}]{cww+08}
Curran S.~J.,  Whiting M.~T.,  Wiklind T.,  Webb J.~K.,  Murphy M.~T.,
  Purcell C.~R.,  2008, MNRAS, 391, 765

\bibitem[\protect\citeauthoryear{{di Serego-Alighieri}, {Danziger}, {Morganti}
  \& {Tadhunter}}{{di Serego-Alighieri} et~al.}{1994}]{ddmt94}
{di Serego-Alighieri} S.,  {Danziger} I.~J.,  {Morganti} R.,    {Tadhunter}
  C.~N.,  1994, MNRAS, 269, 998

\bibitem[\protect\citeauthoryear{{Drinkwater}, {Combes} \&
  {Wiklind}}{{Drinkwater} et~al.}{1996}]{dcw96}
{Drinkwater} M.~J.,  {Combes} F.,    {Wiklind} T.,  1996, A\&A, 312, 771

\bibitem[\protect\citeauthoryear{{Drinkwater}, {Webster}, {Francis}, {Condon},
  {Ellison}, {Jauncey}, {Lovell}, {Peterson} \& {Savage}}{{Drinkwater}
  et~al.}{1997}]{dwf+97}
{Drinkwater} M.~J.,  {Webster} R.~L.,  {Francis} P.~J., et al.,  1997, MNRAS, 284, 85

\bibitem[\protect\citeauthoryear{{Ellison}, {Hall} \& {Lira}}{{Ellison}
  et~al.}{2005}]{ehl05}
{Ellison} S.~L.,  {Hall} P.~B.,    {Lira} P.,  2005, AJ, 130, 1345

\bibitem[\protect\citeauthoryear{Ellison, Yan, Hook, Pettini, Wall \&
  Shaver}{Ellison et~al.}{2001}]{eyh+01}
Ellison S.~L.,  Yan L.,  Hook I.~M.,  Pettini M.,  Wall J.~V.,    Shaver P.,
  2001, A\&A, 379, 393

\bibitem[\protect\citeauthoryear{{Falomo}, {Bersanelli}, {Bouchet} \&
  {Tanzi}}{{Falomo} et~al.}{1993}]{fbbt93}
{Falomo} R.,  {Bersanelli} M.,  {Bouchet} P.,    {Tanzi} E.~G.,  1993, AJ, 106,
  11

\bibitem[\protect\citeauthoryear{{Fiorucci}, {Ciprini} \& {Tosti}}{{Fiorucci}
  et~al.}{2004}]{fct04}
{Fiorucci} M.,  {Ciprini} S.,    {Tosti} G.,  2004, A\&A, 419, 25

\bibitem[\protect\citeauthoryear{{Francis}, {Whiting} \& {Webster}}{{Francis}
  et~al.}{2000}]{fww00}
{Francis} P.~J.,  {Whiting} M.~T.,    {Webster} R.~L.,  2000, PASA, 17, 56

\bibitem[\protect\citeauthoryear{{Fukugita}, {Ichikawa}, {Gunn}, {Doi},
  {Shimasaku} \& {Schneider}}{{Fukugita} et~al.}{1996}]{fig+96}
{Fukugita} M.,  {Ichikawa} T.,  {Gunn} J.~E.,  {Doi} M.,  {Shimasaku} K.,
  {Schneider} D.~P.,  1996, AJ, 111, 1748

\bibitem[\protect\citeauthoryear{{Ghosh}, {Ramsey}, {Sadun} \&
  {Soundararajaperumal}}{{Ghosh} et~al.}{2000}]{grss00}
{Ghosh} K.~K.,  {Ramsey} B.~D.,  {Sadun} A.~C.,    {Soundararajaperumal} S.,
  2000, ApJS, 127, 11

\bibitem[\protect\citeauthoryear{Glikman, Gregg, Lacy, Helfand, Becker \&
  White}{Glikman et~al.}{2004}]{ggl+04}
Glikman E.,  Gregg M.~D.,  Lacy M.,  Helfand D.~J.,  Becker R.~H.,    White
  R.~L.,  2004, ApJ, 607, 60

\bibitem[\protect\citeauthoryear{{Hambly}, {MacGillivray}, {Read}, {Tritton},
  {Thomson}, {Kelly}, {Morgan}, {Smith}, {Driver}, {Williamson}, {Parker},
  {Hawkins}, {Williams} \& {Lawrence}}{{Hambly} et~al.}{2001}]{hmr+01}
{Hambly} N.,  {MacGillivray} H.,  {Read} M.,  et al.,  2001, MNRAS,
  326, 1279

\bibitem[\protect\citeauthoryear{{Isobe}, {Feigelson} \& {Nelson}}{{Isobe}
  et~al.}{1986}]{ifn86}
{Isobe} T.,  {Feigelson} E.,    {Nelson} P.,  1986, ApJ, 306, 490

\bibitem[\protect\citeauthoryear{{Jorgenson}, {Wolfe}, {Prochaska} \&
  {Carswell}}{{Jorgenson} et~al.}{2009}]{jwpc09}
{Jorgenson} R.~A.,  {Wolfe} A.~M.,  {Prochaska} J.~X.,    {Carswell} R.~F.,
  2009, ApJ, 704, 247

\bibitem[\protect\citeauthoryear{{Kanekar}, {Carilli}, {Langston}, {Rocha},
  {Combes}, {Subrahmanyan}, {Stocke}, {Menten}, {Briggs} \&
  {Wiklind}}{{Kanekar} et~al.}{2005}]{kcl+05}
{Kanekar} N.,  {Carilli} C.~L.,  {Langston} G.~I.,  et al.,  2005, PhRvL, 95, 261301

\bibitem[\protect\citeauthoryear{{Lavalley}, {Isobe} \& {Feigelson}}{{Lavalley}
  et~al.}{1992}]{lif92a}
{Lavalley} M.~P.,  {Isobe} T.,    {Feigelson} E.~D.,  1992, in BAAS Vol.~24,
  {ASURV, Pennsylvania State University. Report for the period Jan 1990 - Feb
  1992.}.
pp 839--840

\bibitem[\protect\citeauthoryear{{Liszt} \& {Lucas}}{{Liszt} \&
  {Lucas}}{1996}]{ll96a}
{Liszt} H.,  {Lucas} R.,  1996, A\&A, 314, 917

\bibitem[\protect\citeauthoryear{Murphy, Curran \& Webb}{Murphy
  et~al.}{2003}]{mcw02}
Murphy M.~T.,  Curran S.~J.,    Webb J.~K.,  2003, MNRAS, 342, 830

\bibitem[\protect\citeauthoryear{{Noterdaeme}, {Ledoux}, {Petitjean} \&
  {Srianand}}{{Noterdaeme} et~al.}{2008}]{nlps08}
{Noterdaeme} P.,  {Ledoux} C.,  {Petitjean} P.,    {Srianand} R.,  2008, A\&A,
  481, 327

\bibitem[\protect\citeauthoryear{{Ojha}, {Zacharias}, {Hennessy}, {Gaume} \&
  {Johnston}}{{Ojha} et~al.}{2009}]{ozh+09}
{Ojha} R.,  {Zacharias} N.,  {Hennessy} G.~S.,  {Gaume} R.~A.,    {Johnston}
  K.~J.,  2009, AJ, 138, 845

\bibitem[\protect\citeauthoryear{{Papadakis}, {Samaritakis}, {Boumis} \&
  {Papamastorakis}}{{Papadakis} et~al.}{2004}]{psbp04}
{Papadakis} I.~E.,  {Samaritakis} V.,  {Boumis} P.,    {Papamastorakis} J.,
  2004, A\&A, 426, 437

\bibitem[\protect\citeauthoryear{Pickett, Poynter, Cohen, Delitsky, Pearson \&
  M\"{u}ller}{Pickett et~al.}{1998}]{ppc+98}
Pickett H.~M.,  Poynter R.~L.,  Cohen E.~A.,  Delitsky M.~L.,  Pearson J.~C.,
   M\"{u}ller H. S.~P.,  1998, J. Quant. Spectrosc. Radiat. Transfer, 60, 883

\bibitem[\protect\citeauthoryear{Prochaska, Gawiser, Wolfe, Castro \&
  Djorgovski}{Prochaska et~al.}{2003}]{pgw+03}
Prochaska J.~X.,  Gawiser E.,  Wolfe A.~M.,  Castro S.,    Djorgovski S.~G.,
  2003, ApJ, 595, L9

\bibitem[\protect\citeauthoryear{Prochaska, Herbert-Fort \& Wolfe}{Prochaska
  et~al.}{2005}]{phw05}
Prochaska J.~X.,  Herbert-Fort S.,    Wolfe A.~M.,  2005, ApJ, 635, 123

\bibitem[\protect\citeauthoryear{{Romero}, {Cellone}, {Combi} \&
  {Andruchow}}{{Romero} et~al.}{2002}]{rcca02}
{Romero} G.~E.,  {Cellone} S.~A.,  {Combi} J.~A.,    {Andruchow} I.,  2002,
  A\&A, 390, 431

\bibitem[\protect\citeauthoryear{{Schneider}, {Hall}, {Richards}, {Strauss},
  {Vanden Berk}, {Anderson}, {Brandt}, {Fan} \& {Jester}}{{Schneider}
  et~al.}{2007}]{shr+07}
{Schneider} D.~P.,  {Hall} P.~B.,  {Richards} G.~T.,  et al.,
  2007, AJ, 134, 102

\bibitem[\protect\citeauthoryear{{Skrutskie}, {Cutri}, {Stiening}, {Weinberg},
  {Schneider}, {Carpenter}, {Beichman} \& {Capps}}{{Skrutskie}
  et~al.}{2006}]{scs+06}
{Skrutskie} M.~F.,  {Cutri} R.~M.,  {Stiening} R.,  {Weinberg} M.~D.,
  {Schneider} S.,  {Carpenter} J.~M.,  {Beichman} C.,    {Capps} R. .~M.,
  2006, AJ, 131, 1163

\bibitem[\protect\citeauthoryear{{Srianand}, {Gupta}, {Petitjean}, {Noterdaeme}
  \& {Ledoux}}{{Srianand} et~al.}{2010}]{sgp+10}
{Srianand} R.,  {Gupta} N.,  {Petitjean} P.,  {Noterdaeme} P.,    {Ledoux} C.,
  2010, MNRAS, p.~1888

\bibitem[\protect\citeauthoryear{{Srianand}, {Noterdaeme}, {Ledoux} \&
  {Petitjean}}{{Srianand} et~al.}{2008}]{snlp08}
{Srianand} R.,  {Noterdaeme} P.,  {Ledoux} C.,    {Petitjean} P.,  2008, A\&A,
  482, L39

\bibitem[\protect\citeauthoryear{{Stickel}, {Rieke}, {K\"{u}hr} \&
  {Rieke}}{{Stickel} et~al.}{1996a}]{srkr96}
{Stickel} M.,  {Rieke} G.~H.,  {K\"{u}hr} H.,    {Rieke} M.~J.,  1996a, ApJ,
  468, 556

\bibitem[\protect\citeauthoryear{{Stickel}, {Rieke}, {Rieke} \&
  {K\"{u}hr}}{{Stickel} et~al.}{1996b}]{srrk96}
{Stickel} M.,  {Rieke} M.~J.,  {Rieke} G.~H.,    {K\"{u}hr} H.,  1996b, A\&A,
  306, 49

\bibitem[\protect\citeauthoryear{{Wiklind} \& {Combes}}{{Wiklind} \&
  {Combes}}{1994}]{wc94}
{Wiklind} T.,  {Combes} F.,  1994, A\&A, 286, L9

\bibitem[\protect\citeauthoryear{{Wiklind} \& {Combes}}{{Wiklind} \&
  {Combes}}{1995}]{wc95}
{Wiklind} T.,  {Combes} F.,  1995, A\&A, 299, 382

\bibitem[\protect\citeauthoryear{{Wiklind} \& {Combes}}{{Wiklind} \&
  {Combes}}{1996a}]{wc96}
{Wiklind} T.,  {Combes} F.,  1996a, Nat, 379, 139

\bibitem[\protect\citeauthoryear{{Wiklind} \& {Combes}}{{Wiklind} \&
  {Combes}}{1996b}]{wc96b}
{Wiklind} T.,  {Combes} F.,  1996b, A\&A, 315, 86

\bibitem[\protect\citeauthoryear{{Wiklind} \& {Combes}}{{Wiklind} \&
  {Combes}}{1997}]{wc97}
{Wiklind} T.,  {Combes} F.,  1997, A\&A, 328, 48

\bibitem[\protect\citeauthoryear{{Wiklind} \& {Combes}}{{Wiklind} \&
  {Combes}}{1998}]{wc98}
{Wiklind} T.,  {Combes} F.,  1998, ApJ, 500, 129

\bibitem[\protect\citeauthoryear{{Wright} \& {Otrupcek}}{{Wright} \&
  {Otrupcek}}{1990}]{wo90}
{Wright} A.,  {Otrupcek} R.,  1990, Parkes Catalogue.
Australia Telescope National Facility

\bibitem[\protect\citeauthoryear{Wright, Ables \& Allen}{Wright
  et~al.}{1983}]{waa83}
Wright A.~E.,  Ables J.~G.,    Allen D.~A.,  1983, MNRAS, 205, 793

\bibitem[\protect\citeauthoryear{{Zacharias}, {Monet}, {Levine}, {Urban},
  {Gaume} \& {Wycoff}}{{Zacharias} et~al.}{2004}]{zml+04}
{Zacharias} N.,  {Monet} D.~G.,  {Levine} S.~E.,  {Urban} S.~E.,  {Gaume} R.,
   {Wycoff} G.~L.,  2004, in Bulletin of the American Astronomical Society
  Vol.~36 of Bulletin of the American Astronomical Society, {The Naval
  Observatory Merged Astrometric Dataset (NOMAD)}.
p.~1418

\bibitem[\protect\citeauthoryear{Zwaan \& Prochaska}{Zwaan \&
  Prochaska}{2006}]{zp06}
Zwaan M.~A.,  Prochaska J.~X.,  2006, ApJ, 643, 675

\end{thebibliography}

\bsp

\label{lastpage}

\end{document}